\documentclass[sigconf]{acmart}
\AtBeginDocument{%
  }

\copyrightyear{2025}
\acmYear{2025}
\acmConference[News Futures @ CHI'25]{Make sure to enter the correct
  conference title from your rights confirmation email}{April 26,
  2025}{Yokohama, Japan}

\usepackage{todonotes}
\let\xtodo\todo
\renewcommand{\todo}[1]{\xtodo[inline,color=green!50]{#1}}

\begin{document}

\title{Rethinking News and Media System Design Towards Positive Societal Implications}

\author{Florian Bemmann}
\email{florian.bemmann@ifi.lmu.de}
\orcid{0000-0002-5759-4976}
\affiliation{%
  \institution{LMU Munich}
  \country{Germany}
}

\author{Doruntina Murtezaj}
\orcid{0009-0004-9480-564X}
\email{doruntina.murtezaj@ifi.lmu.de}
\affiliation{%
  \institution{LMU Munich}
  \country{Germany}
}

\renewcommand{\shortauthors}{Bemmann and Murtezaj}

\begin{abstract}
Since this century, the speed, availability, and plethora of online informational content have made it increasingly difficult for humans to keep an overview of real-world situations, build a personal opinion, and sometimes even decide on the truth.
Thereby, personal opinion-making and public discourse became harder - two essential building blocks that keep a democratic society alive.
HCI thus needs to rethink news, information, and social media systems to mitigate such negative effects. Instead of polarising through emotional and extremely framed messages, informational content online should make people think about other opinions and discuss constructively. Instead, through polarization and filter bubble effects, people lose openness and tolerance for the existence of opposing opinions.
In this workshop, we will discuss how we can redesign our information technology for a better societal impact. We will present key takeaways from the social sciences and discuss how we can implement them using recent HCI findings and digital technologies.
\end{abstract}



\maketitle

\section{Introduction}

Since the last few years, our societies have tended to polarize. People feel their governments poorly represent their interests, lose trust in the democratic system and institutions, and an overall misperception spreads on current societal issues~\cite{guess2021pnas,stark2020algorithms}.
One component that fosters this development is how we consume and exchange information online. Modern news and social media portals leverage people’s desire for simple and emotional messages to show them contents that catch them especially well, increasing their platform usage~\cite{seaver2019captivating,luca2015user}. This leads to a distorted perception of reality and enforces one’s opinion filter bubble~\cite{pariser2011filter}, which leads to polarization and distorted public opinion making.
Furthermore, since this century, increased speed, availability, and plethora of informational content have made it difficult for humans to draw personal conclusions and build opinions. Partially contradictory, misleading, or one-sidedly presented facts make it even more difficult to decide on what to regard as the truth.
All this negatively affects the societal discourse and public opinion-making~\cite{zuiderveen2016should}, which are essential to keep a vivid democratic society~\cite{dahl2020democracy}. 

To improve this situation and steer our systems towards more positive social and societal implications, HCI has to rethink the design of how we select, consume, and interact with news. To come up with ecologically valid interface concepts and systemic improvements, it is essential to work interdisciplinarily and accommodate findings of fields such as communication sciences and sociology, which study the human psychological perspective of these aspects. 

In the workshop, we will discuss how we can leverage emerging HCI concepts and technologies to redesign our information technology for a better societal impact. We will present key takeaways from the social sciences and discuss how we can implement them using recent HCI findings and digital technologies.

\section{The Effects of Fastened Information Technology}

Information technology has ever since been the enabler for our modern societal structures. Participatory societal orders rely on the ability of people to inform themselves and build their own opinions. While public participation was not possible many centuries ago, it was enabled by the rise of information technology in the last century. 
Nowadays, information technology is speeding up and growing so much that it is starting to threaten the advances gained again. People are (subconsciously) overwhelmed by the plethora of available information and do hard selecting of many different sources and even different claimed truths. Furthermore, the growing dominance of large platforms disrupts the media business landscape, resulting in fewer large platforms instead of many small and independent news sources. 
Another key characteristic that distinguishes democratic systems and societies from monarchies is the presence of self-control mechanisms. The growing dominance of a few big players among news organizations weakens the news system's self-control mechanisms. 

Algorithms, such as recommendation and summarization systems, increasingly drive modern news platforms. These algorithms inherently foster sensationalized information to maximize user engagement.
Besides such inherently yielding negative effects, this also opens opportunities for abuse, for example, actively leveraging false- and misinformation~\cite{vosoughi2018spread} (see, for example, cases in elections \cite{allcott2017social}). Populists and other harmful political actors may utilize these channels to increase their power and target weaknesses in the democratic system~\cite{engesser2017populism}. 
Furthermore, all these dynamics have recently sped up through AI. AI content generation tools have lowered the barriers to creating and distributing high-quality content, making sophisticated publishing capabilities accessible to virtually anyone.
Further, AI-based conversational interfaces potentially allow one to reach and discuss with many people at scale.

The availability of recent and reliable information, and especially means to compare different points of view and opinions, are essential for a vivid democratic society~\cite{dahl2020democracy}.

\section{Research Vision}

Research must find ways to mitigate the negative societal effects of news and information systems. We need to raise awareness among system designers and users about the influences that fast, many, and omnipresent news contents have on their users. In the following, we outline ideas from the social sciences and how they could be implemented in HCI.

\subsection{The Social Sciences Perspective: What We Need to Achieve}

The social sciences, such as sociology and communication sciences, have long investigated the implications of our information consumption behaviors (e.g., \cite{kumpel_social_2022}), as well as underlying motivations and drivers that make us engage with news contents (e.g., \cite{schmid_motives-2024}). Interdisciplinary cooperation is thus important to steer HCI research towards ecologically valid concepts with real-world implications.

\paragraph{Content Contextualization}
Contextual labeling provides additional information on a news post that is not included but relevant to understand its meaning and big picture \cite{morrow2022emerging}. Through a neutral and non-judgmental presentation, contextual information might be less prone to backfire effects and avoids biases imposed by system developers or used technologies. Putting contents into the light of related facts and information, instead of flagging and judging it, seems to be a promising approach to mitigate the negative effects of one-sided or emotional presentations of news posts.

\paragraph{Supporting a Constructive Societal Discourse}

A constructive political discourse is an essential element. Guidelines for constructive political discourse from psychology research (e.g.,\cite{peters2005public}) differentiate between short-term and long-term positive interdependencies. Short-term aims evolve around reaching an immediate consensus among citizens on a specific course of action to solve a problem. Long-term aims rather aim for systemic improvements and maintaining the health of the democratic system. Research points out the relevance of constructive controversy~\cite{johnson2014constructive}, and investigates the actual effects of various content labeling and contextualization approaches~\cite{morrow2022emerging} on the discourse.

\paragraph{Sparking Openness and Tolerance for Other Opinions}
Work on characteristics of effective democratic discourse, among others, identifies increasing the cohesiveness in the society through the discourse processes~\cite{peters2005public}. This requires openness and tolerance for other opinions and viewpoints; otherwise, a discourse will be rather offensive and non-constructive. 
Skilled disagreement~\cite{johnson2014constructive}, i.e., being open and tolerant of other viewpoints and up for discussing them, although not sharing that opinion, is an essential skill: Disagreeing with someone else’s ideas while respecting their personal competence. This leads to being more interested in learning and more willing to incorporate one’s information and reasoning into their own problem analysis - finally leading to better solutions and consensus~\cite{johnson1991joining}. Without consensus, democratic decision-making does not really work; as with non-consensus, extremely one-sided decisions, the other side's opinion is always unsatisfied and tends not to accept the outcome.

\paragraph{The Organizational Setup of News Systems}
Whereas the landscape of news publishers formerly was diverse and consisted of many different publishers like newspapers, TV- and radio stations, we currently see a tendency towards centralization. Few big players, e.g., news aggregators such as Google News and social media portals like Facebook, X, and YouTube, increasingly pressure the market~\cite{moore2016tech}. Increasing their share in news-related revenues (esp. advertisement revenues) makes it harder for smaller publishers and public service broadcasting to compete. Public service broadcasting (e.g., ARD and ZDF in Germany\footnote{\url{https://www.bundestag.de/resource/blob/547642/7c314e46a0f976ab3afbe9966e6d6889/WD-10-009-18-pdf.pdf}, last accessed 25th of February 2025}) receives among others public fundings, with the mission to provide independent information and engage citizens. In contrast, commercial platforms like Google News and Facebook follow economic interests, thus need to optimize for revenue and user engagement, and are prone to other commercially-driven influences~\cite{rinallo2009does}.

Future news systems should be decoupled from economic interests to avoid the pressure of optimizing for maximum user engagement. We should investigate approaches for decentralized, community-organized platforms instead of giving the power to big global players.

\paragraph{Backfire Effects of Misinformation Flagging}
While HCI is strong in designing and implementing fake-, mis-, and disinformation detection approaches (see the systematic review of \citet{hartwig2024landscape}), questions on their ecologically valid effects remain. Research from the social sciences indicates that interfaces that flag information critically impose backfire effects~\cite{clayton2020real} and that interaction and durability effects should be spent more regard on. \citet{nyhan2021backfire} sees potential in corrective information targeting previously built perceptions and that connections between group-identities and false claims play strong roles.

\subsection{Promising Technological Approaches from HCI}

\paragraph{Augmentation with Contextual Information}
Interface augmentation concepts, such as browser plugins and app overlays, allow users to influence the information presentation of news portals. We propose to augment posts with related information to contextualize them. News content, especially brief headlines and teasers, is often presented from one specific perspective, lacking a classification into the big picture. Additional information that puts a headline's message into context might make its interpretation less polarizing and support the consumer in understanding and reflecting on the situation.

Contextual annotations can be created dynamically in the wild with a large language model and be kept neutral and non-judgmental. This is important to avoid backfire effects (see, e.g., \cite{clayton2020real}) and aligns ethical obligations of not imposing developer or researcher opinions into the content and interface.

While browser plugins and app overlays are rather research tools to study interface changes, the insights obtained should finally inform changes to the platform design itself. We see the integrability of such concepts into existing platforms in the wild. Through augmenting posts instead of downtoning or replacing them, the posts do not lose attractiveness to the audience, whereby our concept is well-applicable in practice without opposing the social media portal’s business interest. As browser plugins are under the user's control, they can be applied in the wild without a platform's cooperation. However, browser plugins as a research tool have a major limitation: Information consumption behavior has shifted to mobile devices mostly (for example, regarding social media, 99\% of engagement now occurs on tablets or smartphones\footnote{\url{https://www.forbes.com/advisor/business/social-media-statistics}, last accessed 19th of March 2025}), but support for mobile browser plugins is rare.

\paragraph{Conversational Agents}

We propose to leverage LLM-based conversational agents (CA) to \textit{(1) contextualize social media and news posts} and \textit{(2) spark openness and tolerance for other opinions}. Through adding neutral, related contextual facts (similar approach as described in the previous paragraph) that illustrate the message of a news post, its polarizing effect on an offensive headline is mitigated. Enhancing the aforementioned static augmentations, conversational interfaces offer a variety of interaction opportunities. Users could approach the system with questions or confront it with their own, maybe opposing opinions. Also, the CA could trigger conversations proactively, exchanging thoughts about other present perspectives and fostering participants' openness to other opinions. Such concepts need to be conceptualized and designed in intensive exchange with researchers from the social and communication sciences.

\paragraph{Public Displays}
Public displays serve as powerful tools for disseminating information, engaging passersby, and fostering public discourse in shared spaces \cite{10.1145/3703465.3703470}. They allow information to be broadcasted to diverse audiences without requiring active user engagement, making them particularly effective for raising awareness on societal issues. In the context of news consumption and societal discourse, such displays can provide contextualized news, highlight multiple perspectives, and encourage critical thinking without the filter bubble effects prevalent in algorithm-driven digital platforms.

Public displays offer significant advantages over personal devices in news dissemination by reaching diverse audiences simultaneously and fostering serendipitous engagement with public-interest news. Unlike personal devices, which rely on individual searches or algorithmic feeds, public displays push information directly into shared environments like transit hubs, libraries, and public squares, ensuring equitable access to timely and relevant news \cite{10.1145/1873951.1874203}.

Additionally, public displays can deliver balanced, locally relevant news tailored to the audience through generative AI and large language models, ensuring content is both accessible and engaging. Interactive public displays further enhance civic participation by prompting discussions, soliciting feedback, or supporting fact-checking initiatives. For example, they can enable citizen engagement through polls, interactive Q\&A formats, or community-sourced content, encouraging reflection on local governance and political events.

\paragraph{Gamified Interfaces}
Gamification elements have been successful in educational contexts and in forming emotional skills \cite{wulansari2020video}, and also been a tool in many behavior-change-related research projects \cite{huber2015gamification}. Their strength is a strong and often inherent user engagement that other concepts often lack. Well-designed gamification elements catch their users and motivate interaction without requiring any further extrinsic motivational component. Game elements can thereby be implemented more or less explicitly; for example, \citet{murnane2020ambient} integrate gamified feedback on individual behavior in existing smartphone UIs. We motivate HCI researchers to revisit such concepts and see how they could be leveraged for societal informational sustainability.

\section{Challenges To Be Discussed in the Workshop}

To advance HCI research on societally sustainable information technology, we will discuss the following aspects. We think that the aspects and ideas presented in this paper can inspire researchers from various fields and motivate further projects.

\begin{itemize}
    \item \textbf{Accommodating interdisciplinary insights} is vital to design ecologically valid solutions. How can we reach a collaboration culture that includes researchers from other non-HCI fields more often?
    \item Ecologically valid interface concepts. While studies often focus on isolated sub-aspects of an interface design, evaluations of real-world effects with systems deployed in the wild are rare. We argue that research concepts should be made with \textbf{real-world applicability} in mind, finally aiming for a field study.
    \item To what extent can we, as HCI researchers, make a difference? Sustainable behavior change research in other domains, such as environmental sustainability, recently considered itself in a dead-end, questioning whether HCI is a field that can actually have an impact~\cite{bremer2022havewe}. We should think about to what extent these considerations also apply to societal sustainability research and be aware of \textbf{naturally-given limitations}.
    \item Deploying \textbf{public displays for news consumption presents unique challenges}. Issues of credibility, content governance, and potential misuse must be carefully considered, particularly regarding misinformation, censorship, or over-commercialization.
    \item The \textbf{ethics of interface design}. When building systems that aim to have some effect on people individually or on a larger scale, the values, ideas, and opinions they convey have to be considered carefully. Incorporating any developer-defined opinion needs to be questioned critically. We do not have the right to claim that our opinion of "what is good for people" actually is what they should accommodate.
\end{itemize}

\bibliographystyle{ACM-Reference-Format}
\bibliography{sample-base}

\end{document}